\documentclass[conference]{IEEEtran}
\IEEEoverridecommandlockouts
\usepackage{amsmath,amssymb,amsfonts}
\usepackage{algorithmic}
\usepackage{graphicx}
\usepackage{textcomp}
\usepackage{xcolor}

\usepackage{threeparttable}
\usepackage[square,sort,comma,numbers]{natbib}
\usepackage{hyperref}

\def\BibTeX{{\rm B\kern-.05em{\sc i\kern-.025em b}\kern-.08em
    T\kern-.1667em\lower.7ex\hbox{E}\kern-.125emX}}

\author{

    \IEEEauthorblockN{1\textsuperscript{st} Te-Lun Yang}
    \IEEEauthorblockA{
        \textit{Graduate Institute of Networking and Multimedia} \\
        \textit{National Taiwan University} \\
        d12944007@csie.ntu.edu.tw
    }
    
    \\
    
    \IEEEauthorblockN{2\textsuperscript{nd} Jyi-Shane Liu}
    \IEEEauthorblockA{
        \textit{Department of Computer Science} \\
        \textit{National Chengchi University} \\
        liujs@nccu.edu.tw
    }
    
    \\
    
    \IEEEauthorblockN{3\textsuperscript{rd} Yuen-Hsien Tseng}
    \IEEEauthorblockA{
        \textit{Graduate Institute of Library \& Information Studies} \\ \textit{National Taiwan Normal University} \\
        samtseng@ntnu.edu.tw
    }
    
    \\
    
    \IEEEauthorblockN{4\textsuperscript{th} Jyh-Shing Roger Jang}
    \IEEEauthorblockA{
        \textit{Department of Computer Science \& Information Engineering} \\
        \textit{National Taiwan University} \\
        jang@mirlab.org
    }
}

\title{Knowledge Retrieval Based on Generative AI}

\begin{document}
\maketitle

\renewcommand{\thefootnote}{\fnsymbol{footnote}}
\footnotetext[1]{To whom all correspondence should be addressed. E-mail: d12944007@csie.ntu.edu.tw}

\begin{abstract}
This study develops a question-answering system based on Retrieval-Augmented Generation (RAG) using Chinese Wikipedia and Lawbank as retrieval sources. Using TTQA and TMMLU+ as evaluation datasets, the system employs BGE-M3 for dense vector retrieval to obtain highly relevant search results and BGE-reranker to reorder these results based on query relevance. The most pertinent retrieval outcomes serve as reference knowledge for a Large Language Model (LLM), enhancing its ability to answer questions and establishing a knowledge retrieval system grounded in generative AI.

The system's effectiveness is assessed through a two-stage evaluation: automatic and assisted performance evaluations. The automatic evaluation calculates accuracy by comparing the model's auto-generated labels with ground truth answers, measuring performance under standardized conditions without human intervention. The assisted performance evaluation involves 20 finance-related multiple-choice questions answered by 20 participants without financial backgrounds. Initially, participants answer independently. Later, they receive system-generated reference information to assist in answering, examining whether the system improves accuracy when assistance is provided.

The main contributions of this research are: (1) Enhanced LLM Capability: By integrating BGE-M3 and BGE-reranker, the system retrieves and reorders highly relevant results, reduces hallucinations, and dynamically accesses authorized or public knowledge sources. (2) Improved Data Privacy: A customized RAG architecture enables local operation of the LLM, eliminating the need to send private data to external servers. This approach enhances data security, reduces reliance on commercial services, lowers operational costs, and mitigates privacy risks.
\end{abstract}


\section{Introduction}
The field of Information Retrieval (IR) has evolved significantly, driven by the need for real-time information through systems like Google and Bing. These systems use keyword-based matching and ranking algorithms to provide relevant results. However, traditional IR systems have limitations, such as low relevance or excessively lengthy outputs when query terms differ from indexed terms. This increases the effort required to find the desired information.

The rise of Generative AI and Large Language Models (LLMs), such as OpenAI's ChatGPT\cite{wu2023brief} and Google's Gemini, offers new ways to enhance IR systems by better understanding user intent and generating natural language responses. Despite these advances, LLMs face challenges, including generating false or outdated information when training data is insufficient or obsolete. Additionally, they may struggle with imbalanced training data, leading to incorrect outputs in specialized domains.

\begin{figure}[htbp]  
    \centering
    \includegraphics[scale=0.4]{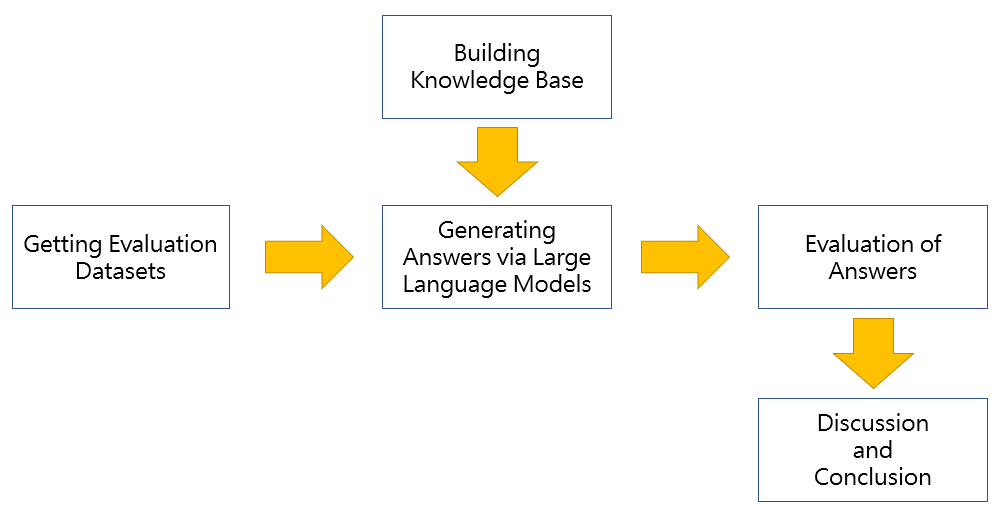}
    \caption{Basic Flow for this study}
    \label{fig:basic_flow_for_this_study}
\end{figure}

To address these issues, the Retrieval-Augmented Generation (RAG) framework combines IR and LLMs, allowing LLMs to retrieve specific information and generate accurate answers without retraining. Dense vector retrieval, a key method in this framework, uses deep learning to map text to high-dimensional vectors, improving accuracy by capturing semantic similarity. This approach is crucial for the development of next-generation knowledge retrieval, overcoming the limitations of traditional methods while leveraging the strengths of LLMs. \autoref{fig:basic_flow_for_this_study} showed basic flow for this study.

\section{Related Work}

\subsection{Information Retrieval}
Information retrieval (IR) has progressed from basic textual data processing to advanced deep learning techniques, with natural language processing (NLP) playing a key role. Early models, such as the Boolean Retrieval Model, used logical operators (OR, AND, NOT) for document-query matching but struggled with complex queries. The Vector Space Model improved upon this by vectorizing documents and queries, using similarity measures, while enhancements like TF-IDF increased retrieval accuracy by adjusting term importance, though lacking contextual understanding. Probabilistic Retrieval Models ranked documents based on relevance probabilities but required large datasets for accuracy. BM25 further refined this approach by dynamically adjusting term weights, and Latent Dirichlet Allocation (LDA) introduced topic modeling using Bayesian networks.
The advent of GPUs and deep learning has popularized Dense Vector Retrieval, which converts text into high-dimensional vectors. Models like BERT and Sentence-BERT leverage these vectors to perform tasks such as sentence classification efficiently, enhancing the semantic understanding of the text and improving IR systems' accuracy and efficiency.

\subsection{Language Model}
\begin{figure}[htbp]  
    \centering
    \includegraphics[scale=0.23]{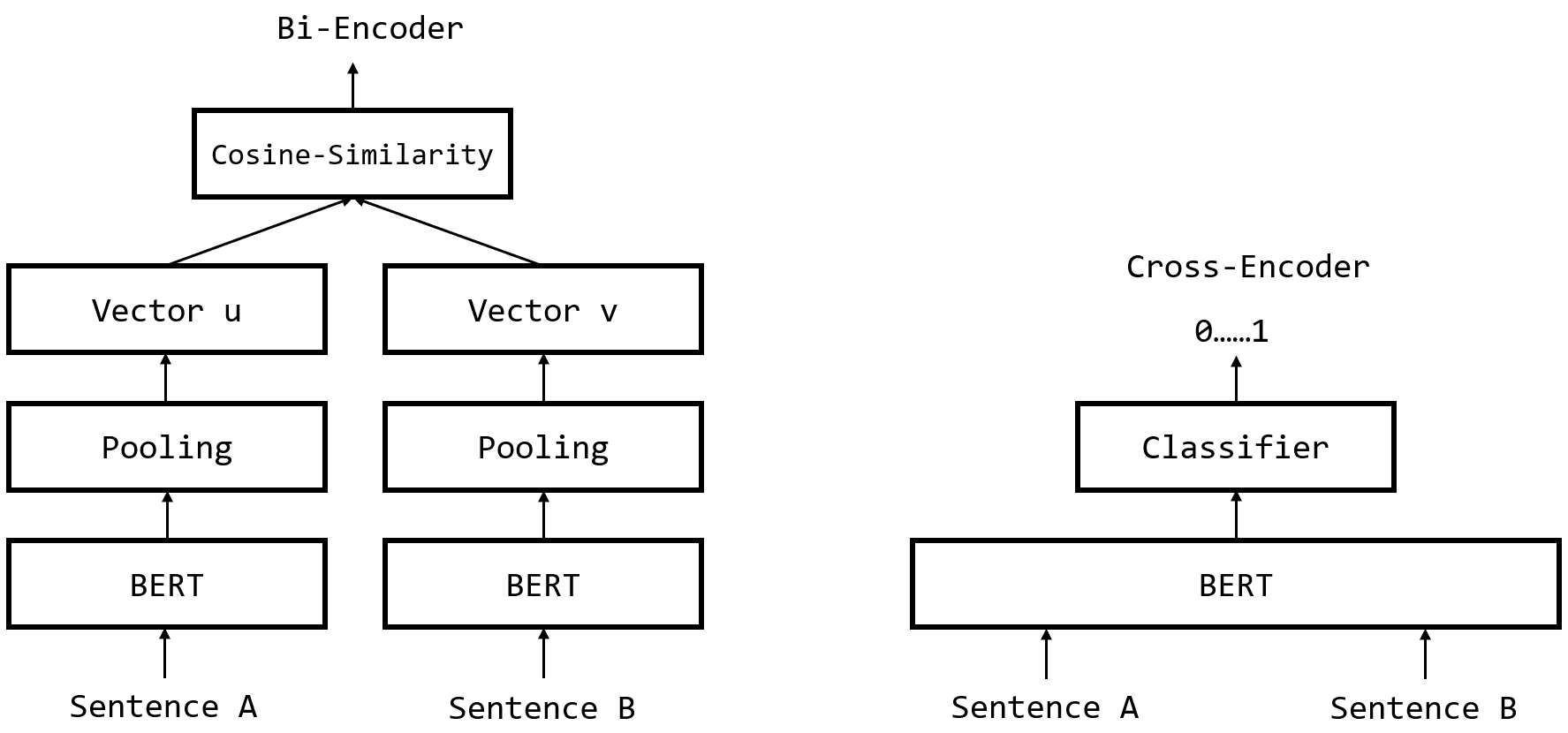}
    \caption{Bi-Encoder and Cross-Encoder}
    \label{fig:bi_encoder_and_cross_encoder}
\end{figure}

Language models (LMs) have progressed from Statistical Language Models (SLMs) to Neural Network Models (NNMs), Pre-trained Models, and Large Language Models (LLMs). Early models like N-grams and Markov Models, though foundational, struggled with scalability due to the "curse of dimensionality." This limitation led to the adoption of NNMs, such as Recurrent Neural Networks (RNNs), which addressed variable-length sequences but faced challenges with long-term dependencies, later mitigated by Long Short-Term Memory (LSTM) and Gated Recurrent Units (GRUs).
The introduction of Distributed Word Representations (DWR), like word2vec, represented a major advancement by mapping words to dense vectors, though it lacked context sensitivity. Pre-trained models, as \autoref{fig:bi_encoder_and_cross_encoder} showed, such as BERT\cite{devlin2018bert} and Sentence-BERT\cite{reimers2019sentence}, resolved this issue with self-attention mechanisms, allowing context-aware embeddings.
The emergence of large language models (LLMs), including GPT-3 and PaLM, has revolutionized NLP by leveraging billions of parameters and exhibiting emergent abilities. These models enable few-shot learning and advance natural language understanding.

\subsection{Retrieval-Augmented Generation}
Retrieval-Augmented Generation (RAG)\cite{lewis2020retrieval} overcomes the limitations of Large Language Models (LLMs), which rely on static, pre-trained datasets that can become outdated and lack domain-specific information. This restricts LLMs' ability to generate accurate and up-to-date responses. RAG integrates Information Retrieval (IR) systems with LLMs, enabling them to query external knowledge sources and access real-time, domain-relevant data. In a typical RAG framework, a retriever processes user queries and retrieves relevant documents based on semantic similarity. These documents are then combined with the original query and passed to the LLM to generate a more accurate and comprehensive response.
RAG also addresses the issue of model obsolescence by dynamically accessing updated information without retraining. However, balancing precision and recall during retrieval can be challenging, and re-ranking is required to prioritize relevant information, ensuring that LLM responses are accurate and contextually appropriate in complex queries.

\subsection{Evaluation}
The rise of Large Language Models (LLMs) necessitates robust evaluation methods across various tasks. Key evaluation metrics include classification, language modeling, text generation, and question answering. For classification, metrics such as precision, recall, and F1-Score are derived from a confusion matrix that assesses true and false positives and negatives.
Language modeling is evaluated using perplexity (PPL), which measures a model's ability to predict token sequences, with lower PPL values indicating stronger performance. Text generation tasks, including machine translation and summarization, use BLEU for precision, analyzing n-gram overlap, and ROUGE for recall, focusing on the longest common subsequences.
Question answering is typically evaluated through accuracy, with Exact Match (EM) and F1-Score commonly applied. For knowledge-based tasks, the MMLU\cite{hendrycks2020measuring} dataset, which contains multiple-choice questions from 57 domains, is used to assess LLMs' generalization and knowledge application capabilities across diverse subjects. These metrics are crucial for evaluating LLM performance across different domains.

\section{Approach}

\subsection{Embedding Model}
The BGE-M3\cite{chen2024bge} model is a highly versatile embedding model supporting multilinguality, multifunctionality, and multi-granularity. Capable of converting text into vectors, it supports over 100 languages for cross-lingual retrieval and is designed to handle sentence, passage, and document-level inputs, with a sequence length of up to 8,192 tokens. It enables dense, sparse, and multi-vector retrieval, making it suitable for a range of tasks from basic keyword matching to complex multi-lingual queries.
BGE-M3’s retrieval process works by embedding the user’s query in one language and retrieving relevant documents in another, supporting both same-language and cross-lingual searches. It is trained on 1.2 billion sentence pairs from 194 languages using unsupervised data, followed by fine-tuning with English and Chinese labeled corpora, and datasets such as MIRACL and Mr.TyDi. The fine-tuning process also involves generating question-answer pairs through GPT-3.5 to create new paired data.
While BGE-M3 excels in vector generation and semantic similarity searches, it may not prioritize relevance in specific tasks such as question answering. This limitation is addressed by BGE-reranker, a cross-encoder designed to re-rank results based on relevance rather than semantic similarity. BGE-reranker assigns relevance scores to the retrieval results generated by BGE-M3, ensuring that higher-ranked results are more likely to accurately answer the query, thus enhancing the effectiveness of retrieval in tasks requiring precise responses.

\subsection{Datasets for Retrieval and Evaluation}
This study uses two primary data sources for retrieval: Chinese Wikipedia, with 1.38 million entries, and Lawbank's fiscal and financial legal regulations, containing 6,193 legal clauses. The Chinese Wikipedia data, stored in dump files and updated monthly, was processed using the BGE-M3 tokenizer, revealing that 1,377,100 articles contained fewer than 8,192 tokens, making them suitable for processing without chunking.
For legal regulations, the analysis showed that out of 151 regulations, 86 contained fewer than 3,400 tokens, with only a few exceeding 17,000 tokens. Due to the specialized nature of legal documents, chunking was adjusted by segmenting the data into individual legal clauses. Among the 6,193 legal clauses, 5,450 contained fewer than 250 tokens, ensuring compatibility with the BGE-M3 model.
The study also employs two evaluation datasets: TC-Eval\cite{hsu2023advancing} and TMMLU+\cite{tamtmmlu+} from iKala. TC-Eval is a Traditional Chinese evaluation suite covering tasks such as contextual question answering, knowledge, and summarization. TTQA (Taiwanese Trivia Question Answering) from TC-Eval contains 103 multiple-choice questions across 15 categories, such as geography and culture. TMMLU+, released by iKala in December 2023, includes 66 topics and 23,015 multiple-choice questions, covering a wide range of professional fields such as law, engineering, and medicine. To ensure consistent testing, the TMMLU+ dataset was reformatted to match the TTQA structure, allowing for efficient comparison and performance evaluation across language models, including handling complex professional-level questions.

\subsection{Vector Index}
Efficient retrieval of high-dimensional vector data, typically generated by embedding models, requires a specialized data structure known as a vector index. This study employs FAISS\cite{douze2024faiss} (Facebook AI Similarity Search), developed by Meta's Fundamental AI Research team, to build vector indices for similarity search and clustering of dense vectors. FAISS supports three common indexing methods: Flat Index for small datasets, providing linear searches across all vectors; Inverted File Index (IVF) for large datasets, which clusters data to reduce search time; and Hierarchical Navigable Small World Graph Index (HNSW), a graph-based method that finds neighboring vectors in high-dimensional spaces with enhanced efficiency.
FAISS also supports three methods for similarity search: L2 Norm, Dot Product, and Cosine Similarity. The L2 Norm measures the Euclidean distance between two vectors and is often used in tasks requiring precise distance calculations. The Dot Product calculates the sum of the products of vector components and is commonly used for similarity and projection tasks in recommendation systems. Cosine Similarity measures the angle between vectors, focusing on direction rather than magnitude, making it popular in information retrieval (IR) systems.
In this experiment, after the BGE-M3 model converts the retrieval data into vectors, FAISS organizes these vectors and constructs the vector index, enabling efficient retrieval based on similarity measures.

\subsection{Large Language Model}
This study utilizes 14 large language models (LLMs) for experimental research, including models such as Taiwan-LLM\cite{lin2023taiwan}, Mixtral 8x7B, Meta llama 2, MediaTek Breeze\cite{hsu2024breeze}, INX Bailong, Google Gemma, ChatGPT 3.5, and Taide. Taiwan LLM, based on llama-2, is fine-tuned for Traditional Chinese, enhancing its performance in local contexts. Mixtral 8x7B employs a Sparse Mixture of Experts (SMoE) architecture to reduce computational costs by activating only a subset of experts during inference. Meta’s llama 2 introduces Grouped-Query Attention (GQA), significantly improving processing speed.
MediaTek’s Breeze model builds on Mistral-7B, adding substantial Chinese data to enhance contextual comprehension. INX Bailong, developed by INX, is a llama-2-based model trained mainly on Traditional Chinese and English texts. Google’s Gemma is a lightweight version of Gemini, emphasizing safety and responsible AI practices, employing RLHF to filter out personal and sensitive data. OpenAI’s ChatGPT 3.5, built on the InstructGPT architecture, leverages RLHF to align with human values, achieving strong performance across various NLP tasks.
Finally, Taide, based on llama-2, integrates local language and cultural elements through continual learning, positioning itself as a reliable model for Traditional Chinese content generation.

\section{Experiments}
\begin{figure}[htbp]  
    \centering
    \includegraphics[scale=0.21]{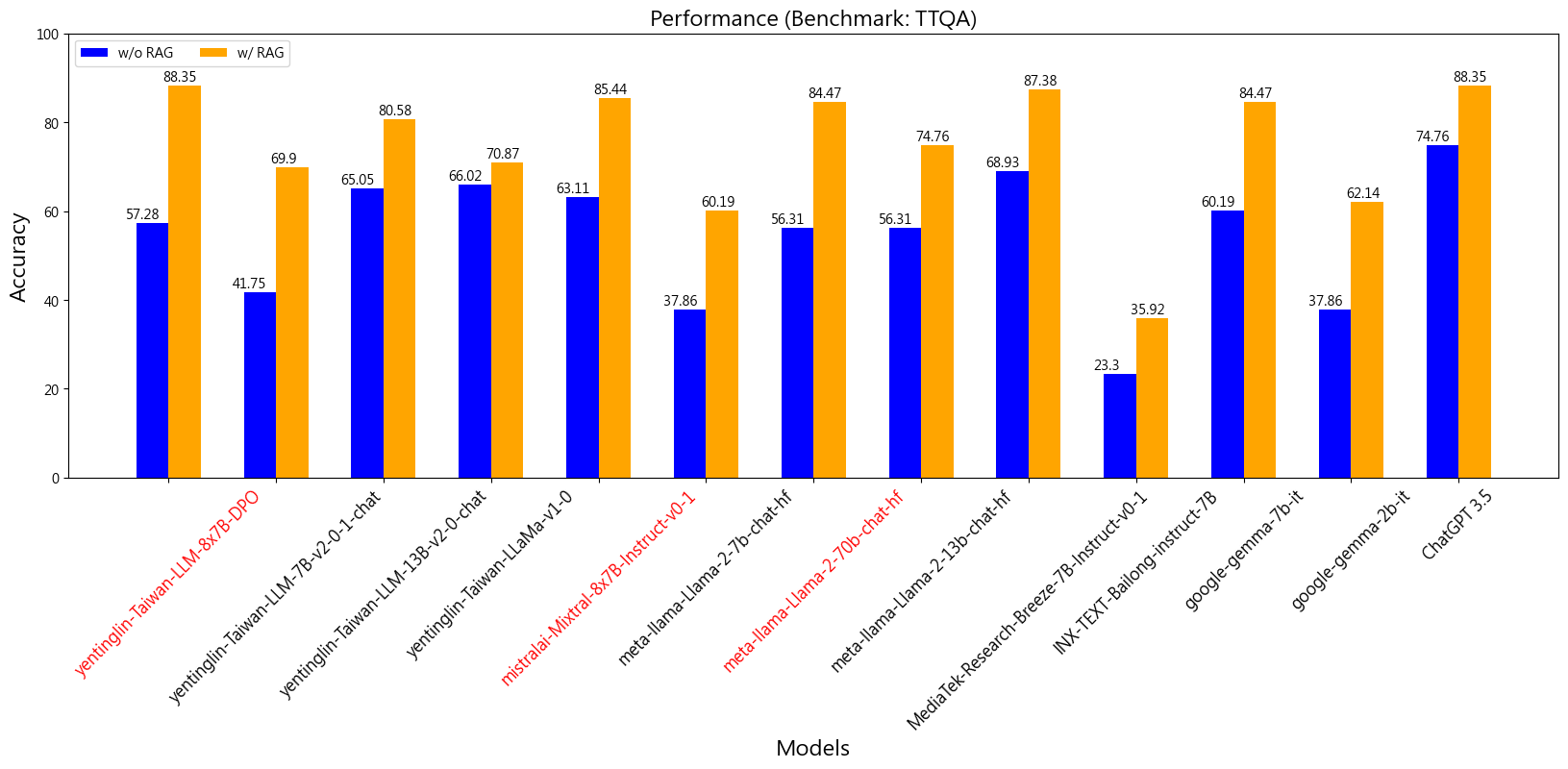}
    \caption{Performance (Benchmark: TTQA)}
    \label{fig:performance_benchmark_ttqa}
\end{figure}

The study evaluated the effectiveness of Retrieval-Augmented Generation (RAG) in improving the accuracy of Large Language Models (LLMs) by answering 103 questions in the TTQA dataset using 14 different models. As \autoref{fig:performance_benchmark_ttqa} showed, the results indicate that models with RAG (w/ RAG) consistently outperformed those without it (w/o RAG), demonstrating the significant impact of RAG on accuracy. Specifically, as \autoref{tab:list_of_scores_for_evaluating_the_ttqa_dataset} showed, Taiwan-LLM-8x7B-DPO saw an accuracy increase from 57.28\% to 88.35\% with RAG, while Taiwan-LLM-7B-v2-0-1-chat and Mistral-8x7B-Instruct-v0-1 improved from 41.75\% to 69.9\% and from 37.86\% to 60.19\%, respectively. ChatGPT 3.5 also showed notable accuracy enhancement from 74.76\% to 88.35\% when supplemented with system-generated reference materials. This trend underscores RAG’s efficacy in enhancing model performance across diverse LLMs.

\begin{table}[htbp]
\caption{List of scores for evaluating the TTQA dataset}
\resizebox{\columnwidth}{!}{
\begin{tabular}{|l|l|l|}
\hline
\textbf{Models}                            & \textbf{w/o RAG} & \textbf{w/ RAG} \\ \hline
Taiwan-LLM-8x7B-DPO                        & 57.28            & 88.35           \\ \hline
Taiwan-LLM-7B-v2-0-1-chat                  & 41.75            & 69.9            \\ \hline
Taiwan-LLM-13B-v2-0-chat                   & 65.05            & 80.58           \\ \hline
Taiwan-LlaMa-v1-0                          & 66.02            & 70.87           \\ \hline
mistralai-Mistral-8x7B-Instruct-v0-1       & 63.11            & 85.44           \\ \hline
meta-llama-Llama-2-7b-chat-hf              & 37.86            & 60.19           \\ \hline
meta-llama-Llama-2-70b-chat-hf             & 56.31            & 84.47           \\ \hline
meta-llama-Llama-2-13b-chat-hf             & 56.31            & 74.76           \\ \hline
MediaTek-Research-Breeze-7B-Instruct-v0-1  & 68.93            & 87.38           \\ \hline
INX-TEXI-Bailong-Instruct-7B               & 23.3             & 35.92           \\ \hline
google-gemma-7b-it                         & 60.19            & 84.47           \\ \hline
google-gemma-2b-it                         & 37.86            & 62.14           \\ \hline
ChatGPT 3.5                                & 74.76            & 88.35           \\ \hline
\end{tabular}
\label{tab:list_of_scores_for_evaluating_the_ttqa_dataset}
}

\begin{tablenotes}
  \small
  \item Note: 8x7B and 70b series used quantization for inference.
\end{tablenotes}

\end{table}

In the evaluation of TMMLU+, covering 66 topics and 23,015 questions, four models were tested: Google’s gemma-7b-it, Meta’s Llama-2-13b-chat-hf, MediaTek’s Breeze-7B-Instruct-v0\_1, and TAIDE-LX-7B-Chat. Performance was assessed with and without the Retrieval-Augmented Generation (RAG) approach. 

\begin{figure}[htbp]  
    \centering
    \includegraphics[scale=0.16]{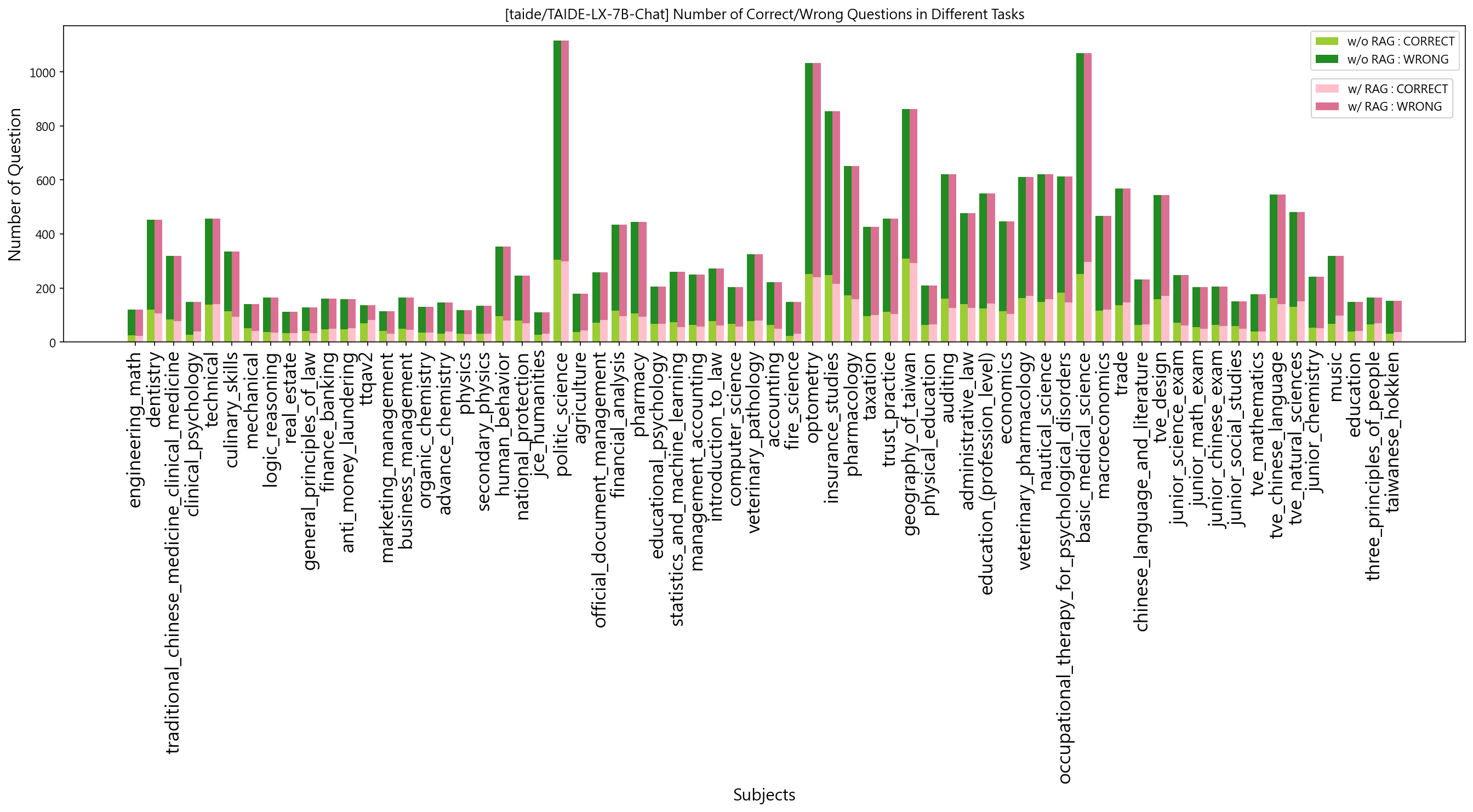}
    \caption{Performance for TAIDE-LX-7B-Chat}
    \label{fig:performance_benchmark_tmmluplus_taide_lx_7b}
\end{figure}

As \autoref{fig:performance_benchmark_tmmluplus_taide_lx_7b} showed, TAIDE-LX-7B-Chat demonstrated decreased accuracy when RAG was applied, performing better as a standalone LLM. 

\begin{figure}[htbp]  
    \centering
    \includegraphics[scale=0.16]{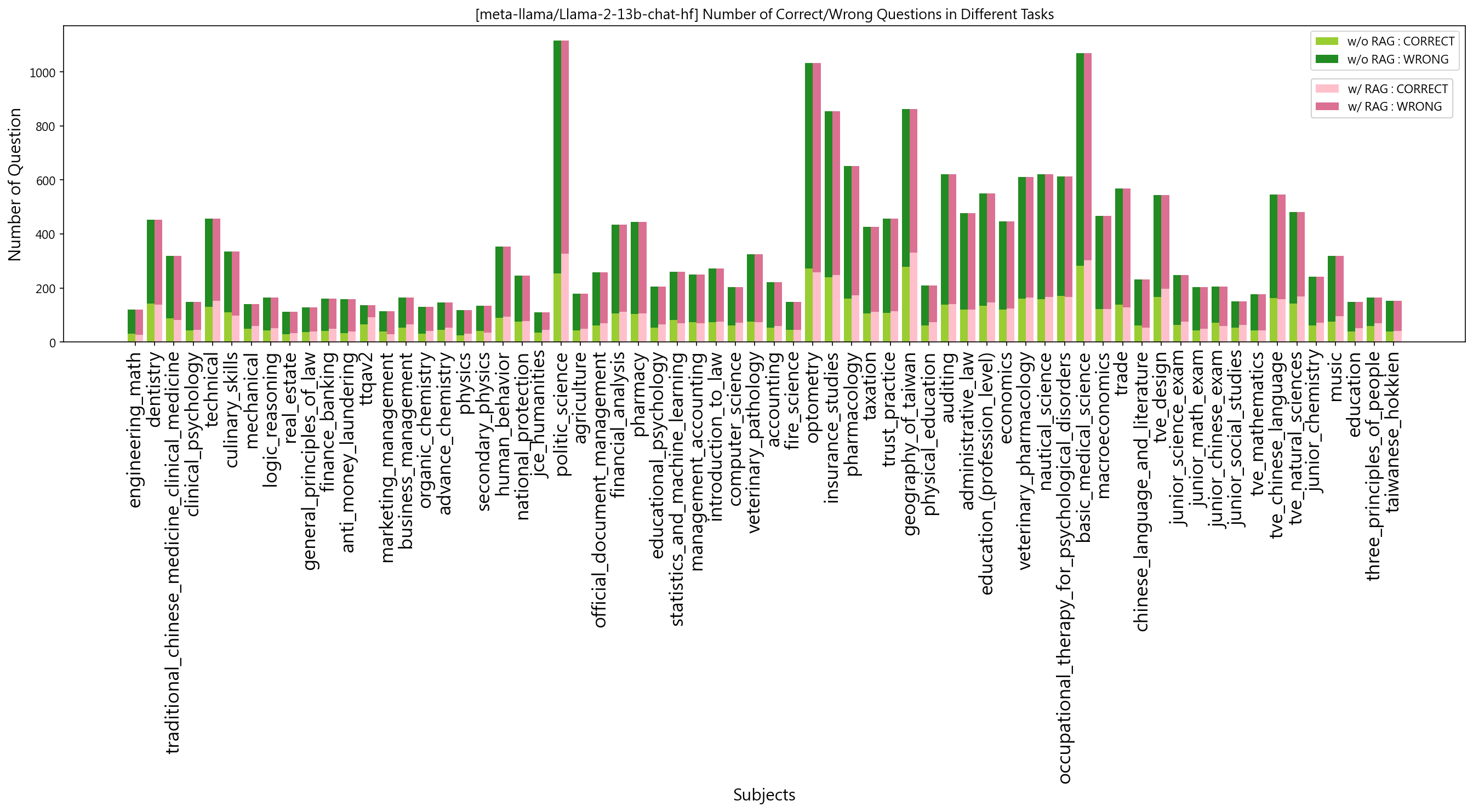}
    \caption{Performance for Llama-2-13b-chat-hf}
    \label{fig:performance_benchmark_tmmluplus_llama2_13b}
\end{figure}

As \autoref{fig:performance_benchmark_tmmluplus_llama2_13b} showed, Llama-2-13b-chat-hf yielded better results than TAIDE-LX-7B-Chat, showing slight accuracy gains in certain topics with RAG, though the majority of results remained similar to the LLM-only configuration. 

\begin{figure}[htbp]  
    \centering
    \includegraphics[scale=0.16]{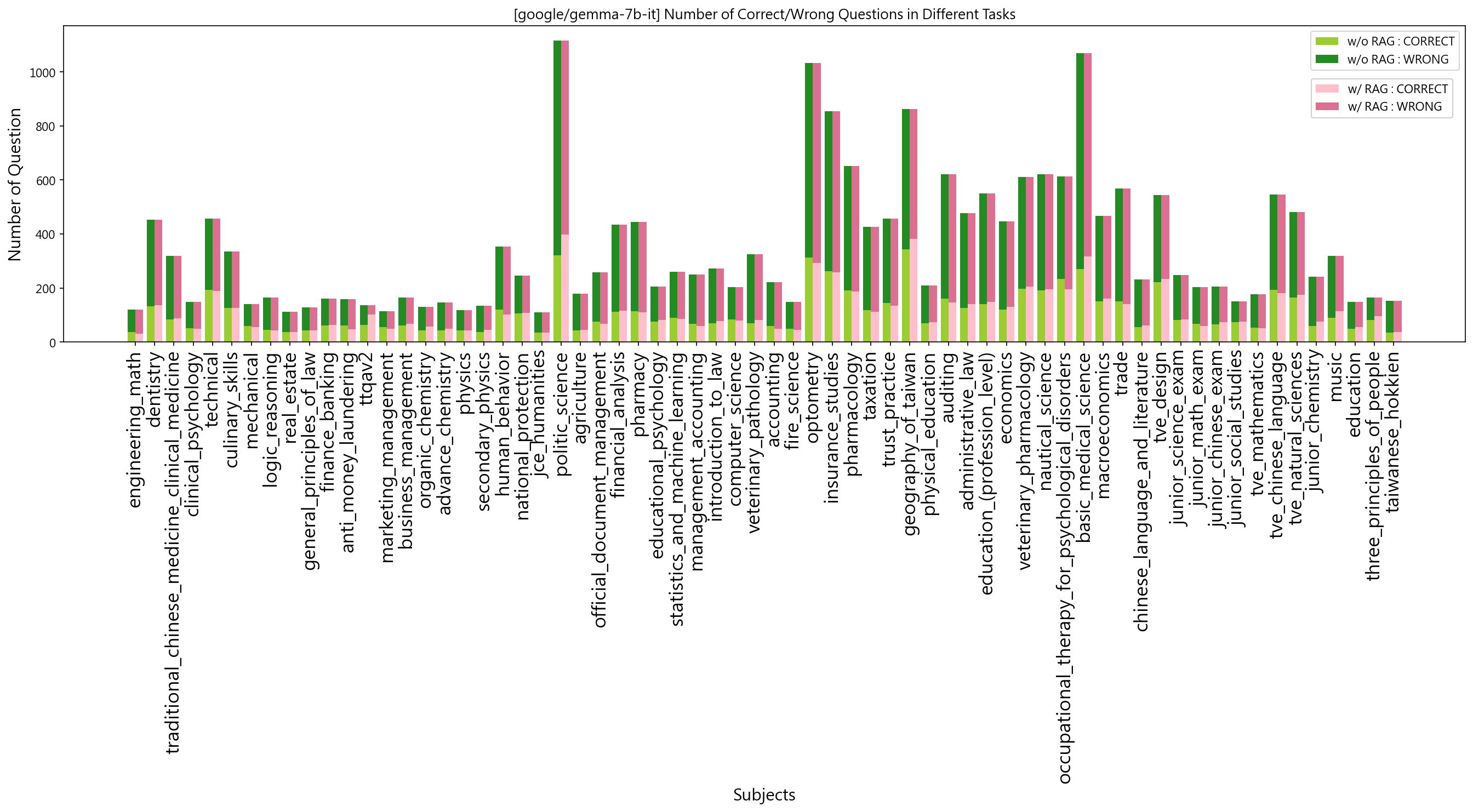}
    \caption{Performance for gemma-7b-it}
    \label{fig:performance_benchmark_tmmluplus_gemma_7b}
\end{figure}

As \autoref{fig:performance_benchmark_tmmluplus_gemma_7b} showed, the gemma-7b-it model performed comparably to Llama-2-13b-chat-hf despite its smaller parameter size, reflecting consistent results across topics. 

\begin{figure}[htbp]  
    \centering
    \includegraphics[scale=0.16]{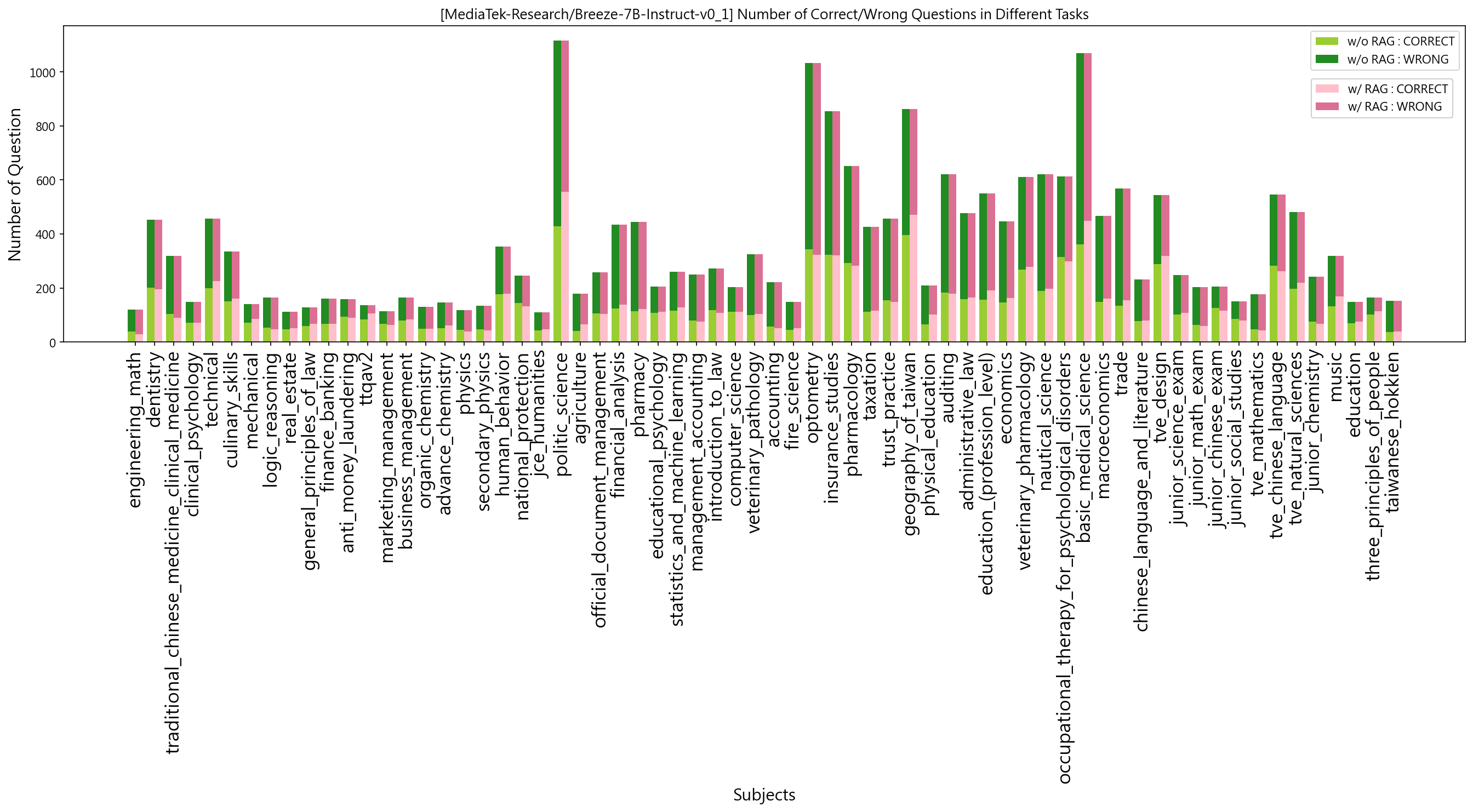}
    \caption{Performance for Breeze-7B-Instruct-v0\_1}
    \label{fig:performance_benchmark_tmmluplus_breeze_7b}
\end{figure}

As \autoref{fig:performance_benchmark_tmmluplus_breeze_7b} showed, Breeze-7B-Instruct-v0\_1 achieved the highest overall accuracy among the models, outperforming others on most topics regardless of RAG usage. For some topics, RAG improved Breeze-7B-Instruct-v0\_1's accuracy, potentially due to this model’s enhanced comprehension of user prompts, underscoring its robustness in both RAG and non-RAG configurations.

Observations from previous experiments indicate that general knowledge sources, such as Chinese Wikipedia, do not adequately support Large Language Models (LLMs) in answering highly specialized questions in fields like engineering mathematics, physics, accounting, and auditing. This limitation contributed to the four LLMs achieving accuracy rates below 50\% across the 66 topics when using the Retrieval-Augmented Generation (RAG) framework. Consequently, the next experiment narrows the TMMLU+ evaluation topics to finance banking and insurance studies. Using legal texts from Lawbank's financial regulations as a specialized retrieval source, we aim to re-evaluate the LLMs' performance, hypothesizing that domain-specific data may improve accuracy.

Using the Retrieval-Augmented Generation (RAG) framework with different retrieval sources revealed notable performance differences among four Large Language Models (LLMs). Without RAG (w/o RAG), models relied solely on their pre-trained capacities, whereas RAG using Chinese Wikipedia (w/ RAG (zhwiki)) and Lawbank's financial regulations (w/ RAG (lawbank)) provided external domain-specific knowledge. 

\begin{figure}[htbp]  
    \centering
    \includegraphics[scale=0.26]{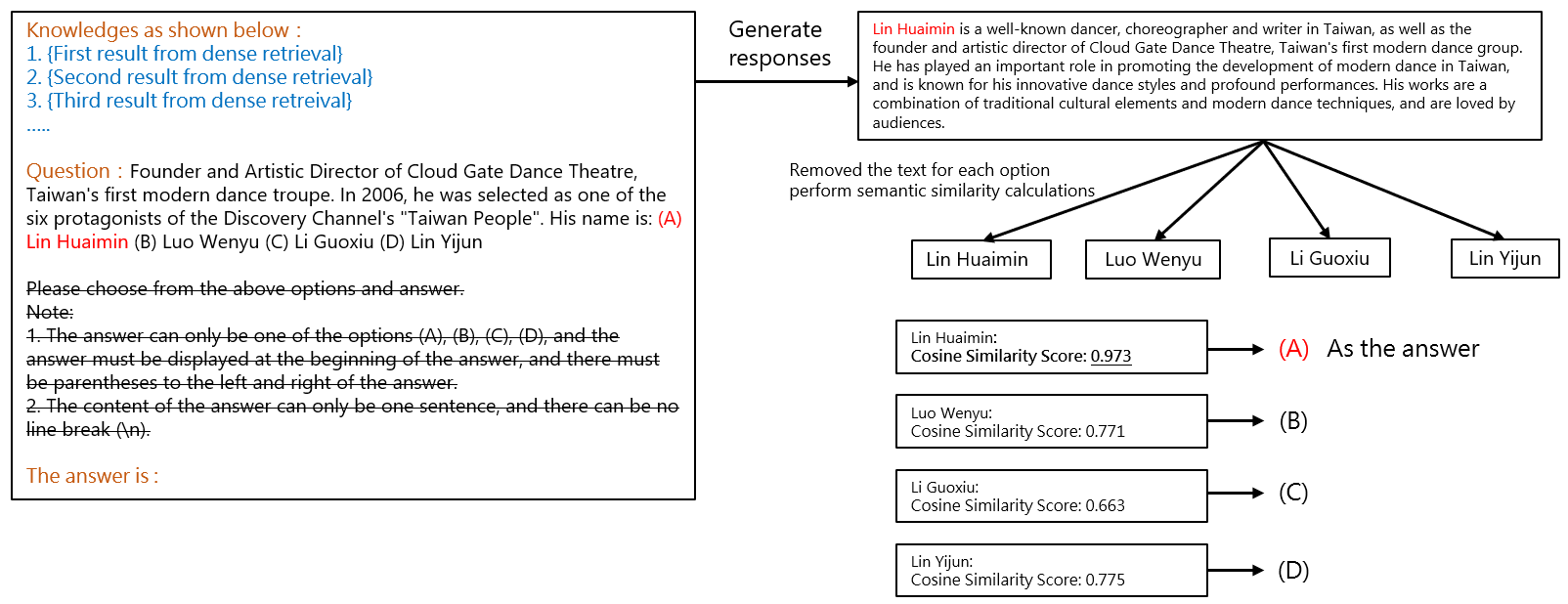}
    \caption{Refine user prompt for multi-choices question}
    \label{fig:refine_user_prompt_for_multichoices_question}
\end{figure}

In \autoref{fig:refine_user_prompt_for_multichoices_question}, results demonstrated that for financial banking questions (160 in total), w/ RAG (lawbank) significantly outperformed w/ RAG (zhwiki), with accuracy improvements of 13.74\% for Gemma-7B-it, 16.25\% for Breeze-7B-Instruct-v0\_1, 5\% for Llama-2-13B-chat-HF, and 3.75\% for TAIDE-LX-7B-Chat. Similarly, for insurance-related questions (855 in total), w/ RAG (lawbank) improved results by 7.25\%, 15.09\%, 2.34\%, and 7.37\%, respectively, across the same models.

Enhancements in LLM response strategies further improved outcomes. Initially, user prompts were revised to allow for unconstrained response generation by removing format restrictions. Additionally, categorical labels such as (A) Lin Huaimin, (B) Luo Wenyu were stripped, leaving only textual options. Each textual option and the LLM-generated answer were transformed into vector representations using the BGE-M3 embedding model, and cosine similarity was calculated. For example, in one evaluation, the similarity scores were 0.973 for Lin Huaimin, 0.771 for Luo Wenyu, 0.663 for Li Guoxiu, and 0.775 for Lin Yijun. With the highest similarity score, "Lin Huaimin" was identified as the correct answer corresponding to label (A), as \autoref{fig:performance_for_financial_banking_questions} and \autoref{fig:performance_for_insurance_related_questions} shown below:

\begin{figure}[htbp]  
    \centering
    \includegraphics[scale=0.33]{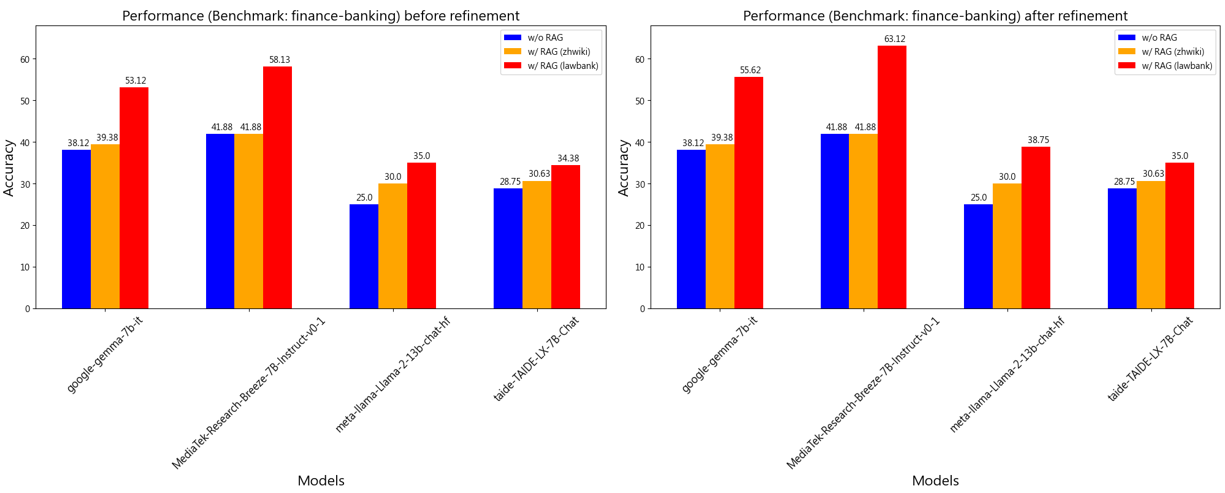}
    \caption{Performance for financial banking questions}
    \label{fig:performance_for_financial_banking_questions}
\end{figure}

\begin{figure}[htbp]  
    \centering
    \includegraphics[scale=0.33]{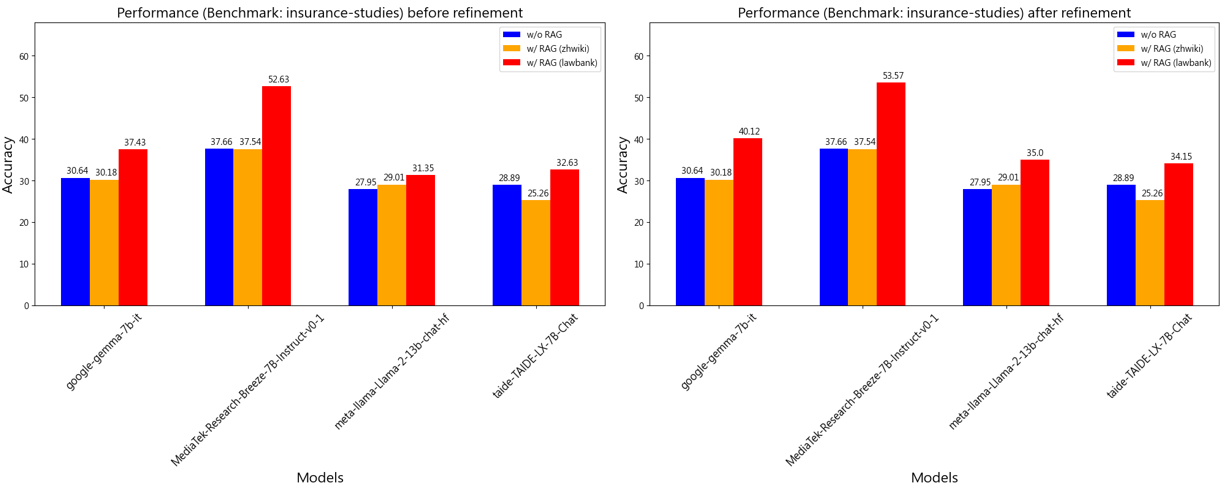}
    \caption{Performance for insurance-related questions}
    \label{fig:performance_for_insurance_related_questions}
\end{figure}

Reevaluating the models post-strategy refinement with w/ RAG (lawbank) demonstrated additional accuracy gains. In financial banking evaluations, improvements were 16.24\% for Gemma-7B-it, 21.24\% for Breeze-7B-Instruct-v0\_1, 8.75\% for Llama-2-13B-chat-HF, and 4.37\% for TAIDE-LX-7B-Chat. For insurance, the respective gains were 9.94\%, 16.03\%, 5.99\%, and 8.89\%. These refinements also yielded performance increases compared to pre-refinement results: in financial banking, improvements were 2.5\% for Gemma-7B-it, 4.99\% for Breeze-7B-Instruct-v0\_1, 3.75\% for Llama-2-13B-chat-HF, and 0.62\% for TAIDE-LX-7B-Chat. In insurance evaluations, these gains were 2.69\%, 0.94\%, 3.65\%, and 1.52\%, respectively.

To evaluate the effectiveness of the Retrieval-Augmented Generation (RAG) framework, a two-stage experiment was conducted involving 20 participants with no financial background. From a dataset of 160 financial banking questions, 20 were randomly selected for the evaluation. In the first stage, participants answered the questions without any reference material. In the second stage, the multiple-choice options (A), (B), (C), and (D) were removed, leaving only the question text. The system generated reference information based on the question and its related retrieval results, providing answers in the form of sentences or paragraphs to assist participants in answering.

\begin{figure}[htbp]  
    \centering
    \includegraphics[scale=0.37]{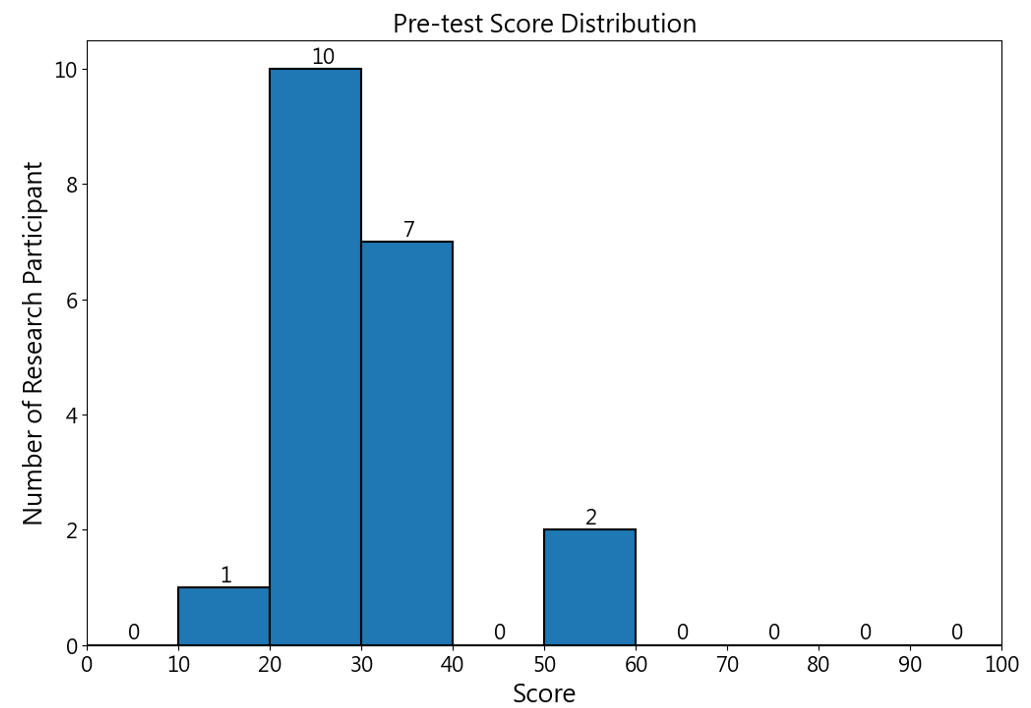}
    \caption{Pre-test score distribution}
    \label{fig:pretest_score_distribution}
\end{figure}

The first stage, as \autoref{fig:pretest_score_distribution} showed, results yielded an average score of 28.75, with a variance of 92.1875, a standard deviation of 9.6014, and both the median and mode at 25. 

\begin{figure}[htbp]  
    \centering
    \includegraphics[scale=0.37]{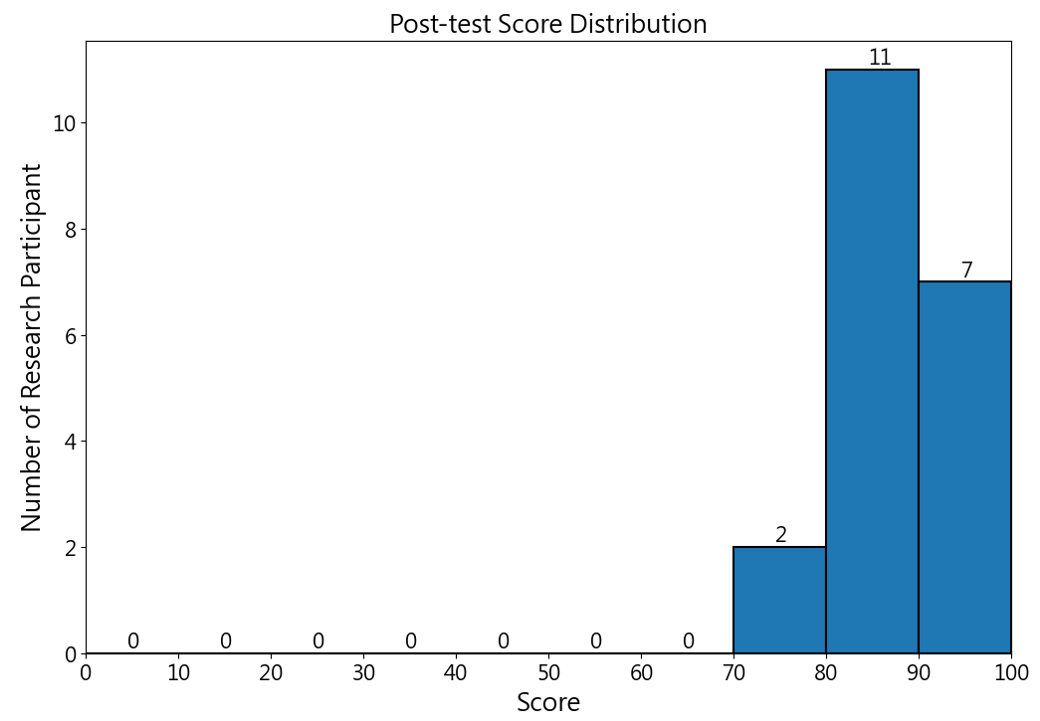}
    \caption{Post-test score distribution}
    \label{fig:posttest_score_distribution}
\end{figure}

In the second stage, as \autoref{fig:posttest_score_distribution} showed, the average score increased significantly to 85.25, with a variance of 36.1875, a standard deviation of 6.0156, and both the median and mode at 85.

This substantial improvement in the mean score between stages indicates that the RAG-generated reference information significantly enhanced the participants' performance.

Moreover, the higher variance and standard deviation in the first stage suggest a wider distribution of scores, reflecting greater variability in participant performance. In contrast, the lower variance and standard deviation in the second stage indicate a more concentrated score distribution, demonstrating more consistent performance across participants. The median and mode in the first stage both being 25 imply that most participants scored around this value, whereas in the second stage, the median and mode shifted to 85, showing that the majority achieved significantly higher scores with the assistance of RAG-generated information.

These results highlight the RAG framework's potential to provide meaningful reference information that can substantially improve performance, particularly for individuals with no prior expertise in the domain. The data underscore the framework's utility in narrowing performance gaps and fostering consistent accuracy in specialized question-answering tasks.

\section{Discussion}
This study successfully developed a knowledge retrieval system based on Retrieval-Augmented Generation (RAG) and evaluated its performance using Chinese Wikipedia and Lawbank as data sources. The experiments conducted with evaluation datasets such as TTQA and TMMLU+ demonstrated that the combination of BGE-M3 and BGE-reranker effectively enhanced the relevance of retrieved results, thereby improving the accuracy of Large Language Models (LLMs) in answering questions. This section discusses the implications of the findings, the practical applications of the system, its limitations, and directions for future research.

\subsection{Interpretation of Findings}
The experimental results reveal that the system exhibits strong performance in the financial and legal domains, particularly in human-assisted question-answering scenarios, where it significantly improved participant accuracy. These findings suggest that incorporating domain-specific knowledge sources, coupled with robust retrieval and ranking mechanisms, can address the knowledge gaps in LLMs. Furthermore, the integration of dense vector retrieval and re-ranking techniques substantially mitigated hallucination effects and enhanced the reliability of the system's responses. These results validate the utility of the RAG framework in addressing knowledge-intensive tasks and highlight the critical role of retrieval result ranking in boosting LLM performance.

\subsection{Comparison with Existing Research}
Compared to traditional information retrieval methods such as TF-IDF and BM25, this study leveraged dense vector retrieval with a multilingual embedding model (BGE-M3) to achieve superior precision. Additionally, the re-ranking model (BGE-reranker) further improved the relevance of retrieved results, enabling the RAG framework to deliver more accurate responses to domain-specific queries. Unlike systems that rely solely on LLMs to generate answers, the proposed system not only enhances accuracy but also reduces reliance on outdated or potentially erroneous information.

\subsection{Practical Implications}
The proposed knowledge retrieval system has significant potential for applications across multiple domains. In the legal sector, it can support decision-making by quickly retrieving relevant statutes and providing concise explanations to assist legal professionals in handling complex cases. In education, the system can serve as a domain-specific question-answering tool to help students efficiently acquire knowledge. Moreover, the locally deployable RAG framework ensures data privacy and security, reducing dependency on commercial cloud services, and thus provides a model for building enterprise-level knowledge management systems.

\subsection{Limitations}
Despite the promising outcomes, this study acknowledges several limitations. First, while the data sources (Chinese Wikipedia and Lawbank) are extensive, they may not encompass all necessary domain-specific knowledge, potentially limiting the system's performance on highly specialized queries. Second, the system exhibited limitations in handling computational and logic-based reasoning questions, reflecting the inherent constraints of LLMs in processing such tasks. Third, the computational demands of the BGE-reranker model present challenges for real-time large-scale applications, highlighting the issue of limited computational resources.

\subsection{Future Directions}
To address the aforementioned limitations, several avenues for future research are proposed. First, expanding the scope of data sources by incorporating additional specialized knowledge bases or dynamically updating datasets could enhance the system's adaptability to diverse queries. Second, integrating methods such as Graph Neural Networks (GNNs) or multi-modal data could further improve the system's reasoning capabilities and versatility. Third, optimizing the efficiency of the re-ranking model, or exploring alternative solutions like distillation techniques, could alleviate resource constraints in real-time applications. Lastly, user experience studies should be conducted to refine the system's interface and interaction processes, ensuring effective deployment in various scenarios.

\section{Conclusion}
This study demonstrates the feasibility of the RAG framework in enhancing the accuracy and utility of knowledge retrieval systems, offering a valuable reference for the integration of IR and LLM technologies. Despite its limitations, the findings highlight the potential of the system in domain-specific knowledge retrieval and question-answering tasks. Future optimizations and extensions of this research will further advance the system's performance and broaden its applicability. There are two significant contributions:
\begin{itemize}
\item Enhancing the Capability of Large Language Models (LLMs) in Knowledge-Intensive Tasks: The integration of the BGE-M3 embedding model facilitates dense vector retrieval, enabling the extraction of highly relevant results based on the semantic similarity between queries and retrieved data. These results serve as critical reference knowledge sources for LLMs, thereby improving their accuracy in addressing complex queries. Additionally, the incorporation of the BGE-reranker re-ranking model refines the retrieved results by identifying information that is most relevant to the queries, enhancing the reliability of the generated answers. This approach effectively mitigates hallucination effects and dynamically sources the latest authorized or publicly available knowledge to meet task-specific requirements.
\item Ensuring Data Privacy and Reducing Dependence on Commercial Services: By employing a customized RAG architecture, the proposed system operates entirely on local infrastructure, eliminating the need to transmit private data to external commercial service providers. This design significantly enhances data security and privacy protection. Furthermore, it minimizes reliance on commercial services, thereby reducing potential operational costs and risks of privacy breaches.
\end{itemize}

\bibliography{main}
\bibliographystyle{IEEEtranN}

\vspace{12pt}

\end{document}